\documentclass[prl,twocolumn,showpacs,superscriptaddress]{revtex4}

\usepackage{epstopdf}
\usepackage{amsbsy}
\usepackage{amsmath}
\usepackage{amsfonts}
\usepackage{amsthm}
\usepackage{color}
\usepackage{mathrsfs}
\usepackage{graphicx}
\usepackage{dcolumn}
\usepackage{bm}

\begin{document}

\title{Conditional detection of pure quantum states of light after storage in a waveguide}

\author{Erhan~Saglamyurek}
\thanks{These authors contributed equally.}
\affiliation{Institute for Quantum Information Science, and Department of Physics \& Astronomy, University of Calgary, 2500 University Drive NW, Calgary, Alberta T2N 1N4, Canada}
\author{Neil~Sinclair}
\thanks{These authors contributed equally.}
\affiliation{Institute for Quantum Information Science, and Department of Physics \& Astronomy, University of Calgary, 2500 University Drive NW, Calgary, Alberta T2N 1N4, Canada}
\author{Jeongwan~Jin}
\thanks{These authors contributed equally.}
\affiliation{Institute for Quantum Information Science, and Department of Physics \& Astronomy, University of Calgary, 2500 University Drive NW, Calgary, Alberta T2N 1N4, Canada}
\author{Joshua~A.~Slater}
\thanks{These authors contributed equally.}
\affiliation{Institute for Quantum Information Science, and Department of Physics \& Astronomy, University of Calgary, 2500 University Drive NW, Calgary, Alberta T2N 1N4, Canada}
\author{Daniel~Oblak}
\affiliation{Institute for Quantum Information Science, and Department of Physics \& Astronomy, University of Calgary, 2500 University Drive NW, Calgary, Alberta T2N 1N4, Canada}
\author{F\'{e}lix~Bussi\`{e}res}
\altaffiliation[Current address: ]{GAP-Optique, Univ. of Geneva, Rue de l'\'{E}cole-de-m\'{e}decine 20, 1211 Geneva , Switzerland}
\affiliation{Institute for Quantum Information Science, and Department of Physics \& Astronomy, University of Calgary, 2500 University Drive NW, 
Calgary, Alberta T2N 1N4, Canada}
\author{Mathew~George}
\affiliation{Department of Physics - Applied Physics, University of Paderborn, Warburger Str. 100, 33095 Paderborn, Germany}
\author{Raimund~Ricken}
\affiliation{Department of Physics - Applied Physics, University of Paderborn, Warburger Str. 100, 33095 Paderborn, Germany}
\author{Wolfgang~Sohler}
\affiliation{Department of Physics - Applied Physics, University of Paderborn, Warburger Str. 100, 33095 Paderborn, Germany}
\author{Wolfgang~Tittel}
\affiliation{Institute for Quantum Information Science, and Department of Physics \& Astronomy, University of Calgary, 2500 University Drive NW, Calgary, Alberta T2N 1N4, Canada}

\begin{abstract}

 Conditional detection is an important tool to extract weak signals from a noisy background and is closely linked to heralding, which is an essential component of protocols for long distance quantum communication and distributed quantum information processing in quantum networks.
Here we demonstrate the conditional detection of time-bin qubits after storage in and retrieval from a photon-echo based waveguide quantum memory. Each qubit is encoded into one member of a photon-pair produced via spontaneous parametric down conversion, and the conditioning is achieved by the detection of the other member of the pair. 
Performing projection measurements with the stored and retrieved photons onto different bases we obtain an average storage fidelity of $0.885\pm0.020$, which exceeds the relevant classical bounds and shows the suitability of our integrated light-matter interface for future applications of quantum information processing.

 \end{abstract}



\maketitle


Quantum memories are key elements for future applications of quantum information science such as long-distance quantum communication via quantum repeaters~\cite{briegel1998a,sangouard2011a} and, more generally, distributed quantum information processing in quantum networks~\cite{kimble2008a}. They enable reversible mapping of arbitrary quantum states between travelling and stationary carriers (i.e. light and matter). This reduces the impact of loss on the time required to establish entanglement between distant locations~\cite{briegel1998a}, and allows the implementation of local quantum computers based on linear optics \cite{kok2007a}. However, towards these ends, the successful transfer of a quantum state into the memory must be announced by a heralding signal. When using an individual absorber, such a signal can be derived through the detection of a change of atomic level population~\cite{specht2011a}. In atomic ensembles, this approach is infeasible. Instead, storage is derived from the detection of a second photon that either indicates the absorption~\cite{tanji2009a}, or the presence of the first at the input of the memory~\cite{tittel2010a} (the first approach relies on spontaneous Raman scattering, the second on using pairs of photons). Furthermore, quantum memories must allow on-demand read-out after second-long storage with high efficiency~\cite{tittel2010a,lvovsky2009a}, and, for viable quantum technology, should be robust and simple to operate (e.g. be based on integrated optics).

A lot of progress towards these (and other) figures of merit has been reported over the past few years, including work that explores electromagnetically induced transparency (EIT), as well as photon-echo and cavity QED-based approaches (see \cite{tittel2010a,lvovsky2009a,sangouard2011a,hedges2010a,clausen2010a,saglamyurek2010a,specht2011a,hosseini2011a,zhang2011a,reim2011a} for reviews and latest achievements). Yet, strictly, most of these experiments did not report true heralding -- either heralding was not actually implemented, or the `heralding' signal was generated only after the stored photon left the memory, or the signal could, due to technical issues, only be derived once the stored photon was detected. Nevertheless, experiments that employ photon pairs~\cite{akiba2009a,clausen2010a,saglamyurek2010a,zhang2011a} do gain from conditioning the detection of the stored photon on that of the auxiliary photon (i.e. \textsl{a posteriori} `heralding'): by reducing the effects of loss and detector noise conditioning generally increases the fidelity between the quantum state of the original and the retrieved photon.

Supplementing the experiments on storage of entangled photons~\cite{saglamyurek2010a,akiba2009a,clausen2010a,zhang2011a}, we now report another step towards the goal of building universal, viable, and heralded quantum memory devices -- the storage of photons in pure quantum states in a solid state waveguide, their retrieval, and their conditional detection by means of temporal correlations with auxiliary photons.
We point out that the step to true heralding is minor and of purely technical nature; it simply requires using different, existing, single-photon detectors (see, e.g.,~\cite{eisaman2011a,idquantique2011a}).

\begin{figure}
\begin{center}
\includegraphics[width=\columnwidth]{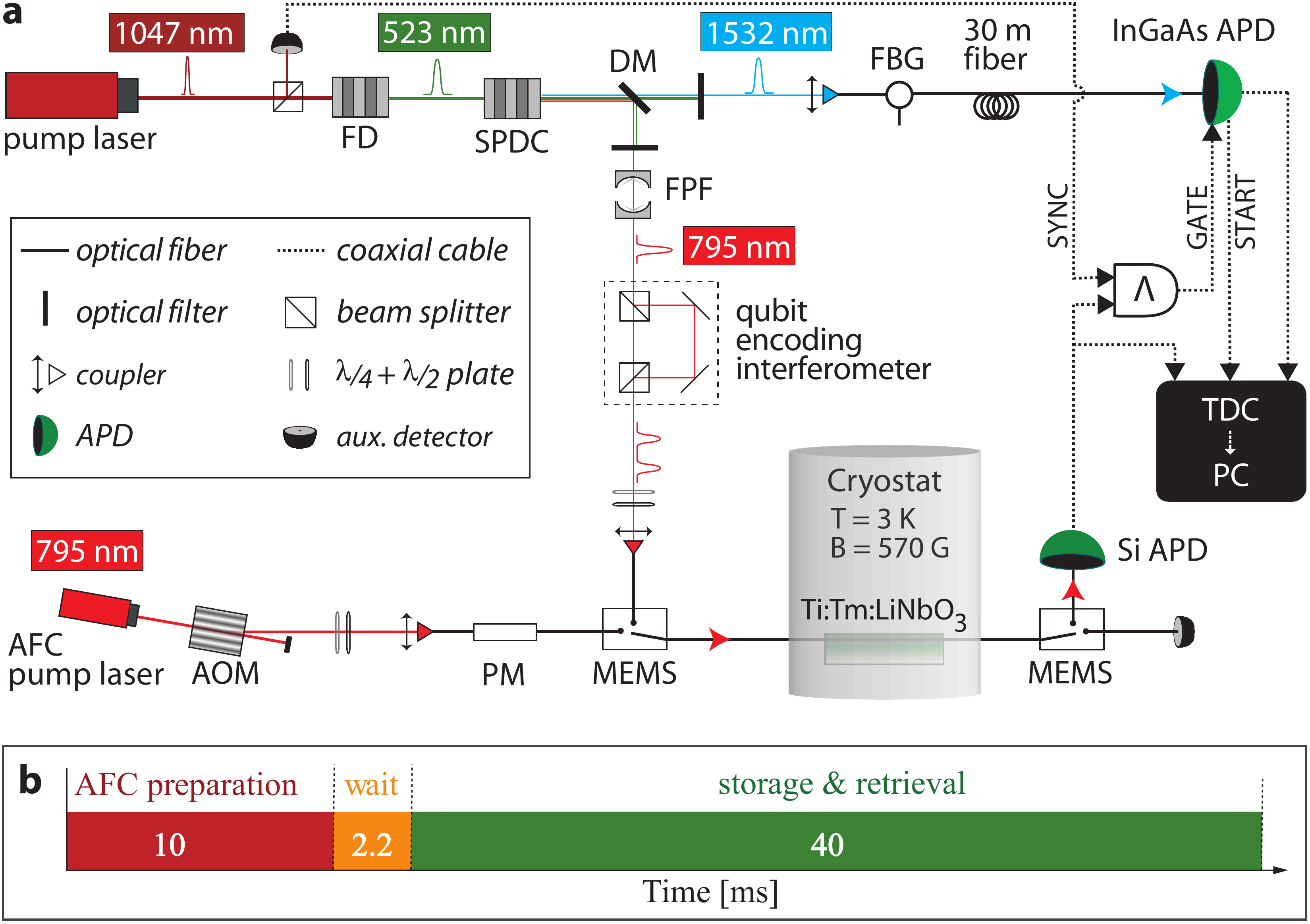}
 \caption{(color online) \textbf{a.} Photon pair source and quantum memory setup (see text for details). Wave-plates align light polarization along the LiNbO$_3$'s C$_3$-axis. The waveguide is held at 3~K, and a 570~G magnetic field is applied along the crystal's C$_3$-axis (see Fig. 2a). \textbf{b.} Timing sequence containing three repeated phases: 10~ms \emph{AFC preparation} for optical pumping, 2.2~ms \emph{wait} to allow excited population to decay, and 40~ms \emph{storage and retrieval}, during which 795~nm photons are successively stored for $t_{st}=6$~ns and then recalled.}
  \label{fig:setup}
  \end{center}
\end{figure}

Our experimental setup consists of two main blocks, see Fig.~\ref{fig:setup}: A spontaneous parametric down-conversion (SPDC) photon-pair source, and a 
Ti:Tm:LiNbO$_3$ single mode waveguide fabricated by indiffusion processes~\cite{sinclair2010a}. When cooled to 3~K, and using a photon-echo quantum memory protocol~\cite{afzelius2009a,tittel2010a,lvovsky2009a}, the Tm-doped waveguide allows storage and retrieval of quantum states encoded into one member of each photon pair, while the detection of the other member provides the conditioning signal. 

In the photon-pair source a mode-locked pump laser generates 6~ps long pulses at a rate of 80~MHz and 1047.328~nm central wavelength. They are subsequently frequency-doubled (FD) in a periodically poled LiNbO$_3$ (PPLN) crystal, yielding pulses with 523.664~nm central wavelength, 16~ps duration, and 90~mW average power. The FD pulses are sent to a second PPLN crystal that, via SPDC, produces pairs of photons centred at 795.506~nm and 1532.426~nm. Frequency filtering the 795~nm photons with a 6~GHz-bandwidth Fabry-Perot filter (FPF) and the 1532~nm photons with a 9~GHz-bandwidth fiber-Bragg grating (FBG) we obtain frequency uncorrelated pairs. Each 795~nm photon travels through an imbalanced Mach-Zehnder interferometer with 42~cm path-length difference, corresponding to 1.4~ns relative delay. Thus, each  photon emerges in a superposition of two temporal modes (early and late), i.e., in a time-bin qubit state~\cite{tittel2001a}. They are then directed into the quantum memory, stored, retrieved, and finally detected by a Si avalanche-photo-diode (APD)-based single-photon detector.

All 1532~nm photons are sent through 30~m standard telecommunication fiber to  an InGaAs APD-based single-photon detector. As is typically done, the detector is gated to reduce noise. The gate signal could in principle be the SYNC signal derived from each pulse emitted by the pump laser. However, as its repetition rate of 80 MHz by far exceeds the maximum gate frequency of our detector, around 1 MHz, we first AND the SYNC pulses with pulses generated by each Si-APD detection, and then use this low-rate signal to gate the InGaAs-APD.  Provided the latter is ready for photon detection (i.e. not deadtime-blocked due to a previous detection), this signal also starts a time-to-digital converter (TDC), which then records the time-difference between the detection events produced by the Si-APD and the InGaAs-APD. This data is used to obtain statistics for single detections of the retrieved 795~nm photons, as well as for detections conditioned on the existence of 1532~nm photons.  
We emphasize that if an InGaAs APD supporting 80 MHz gate rate had been available ~\cite{eisaman2011a,idquantique2011a}, then 1532 nm photons could have been detected without the need for \textsl{a priori} detection of a 795 nm photon. This simple modification of our setup would have turned the conditional detection of 795 nm photons into detections that are heralded by clicks of the InGaAs APD. 

The other main block of our setup is a Ti:Tm:LiNbO$_3$ waveguide  that allows storage and retrieval of the 795~nm photons via the atomic frequency comb (AFC) quantum memory protocol~\cite{afzelius2009a}. This approach to quantum state storage requires the spectral absorption of an atomic ensemble to be constituted of a series of equally spaced lines  with frequency spacing $\Delta_\nu$. The interaction between such an AFC and a photon with wavevector $k$ leads to the absorption of the photon and generates a collective excitation in the atomic medium that is described by
\begin{equation}
	\label{eq:afcqstate}
	\left| \Psi  \right\rangle =\frac{1}{\sqrt{N!}}\sum _{j=1}^{N}\! c_{j} e^{i2\pi m_j\Delta_\nu t} e^{-ikz_j}  \left| g_{1} ,\cdots e_{j} ,\cdots g_{N}  \right\rangle.	
\end{equation}
Here, ${\left| g_{j}  \right\rangle}$ (${\left| e_{j}  \right\rangle}$) denotes the ground (excited) state of atom $j$, $m_j\Delta_\nu$ is the detuning of the atom's transition frequency from the photon carrier frequency, $z_j$ its  position measured along the propagation direction of the light, and the factor $c_j$ depends on the atom's resonance frequency and position. Due to the presence of different atomic transition frequencies, the excited collective coherence dephases rapidly. However, the particular shape of the absorption line results in the recovery of the collective coherence after storage time $t_{st}=1/\Delta_\nu$. This can easily be seen from Eq.~\eqref{eq:afcqstate}: for $t=1/\Delta_\nu$ all frequency dependent phase factors are zero (mod $2\pi$). This leads to re-emission of the photon into the original mode and quantum state with maximally 54\% efficiency for an optimally implemented AFC. Modifications to the procedure enable recall on demand and up to 100\% efficiency~\cite{afzelius2009a}.

\begin{figure}[tt]
\begin{center}
\includegraphics[width=0.95\columnwidth]{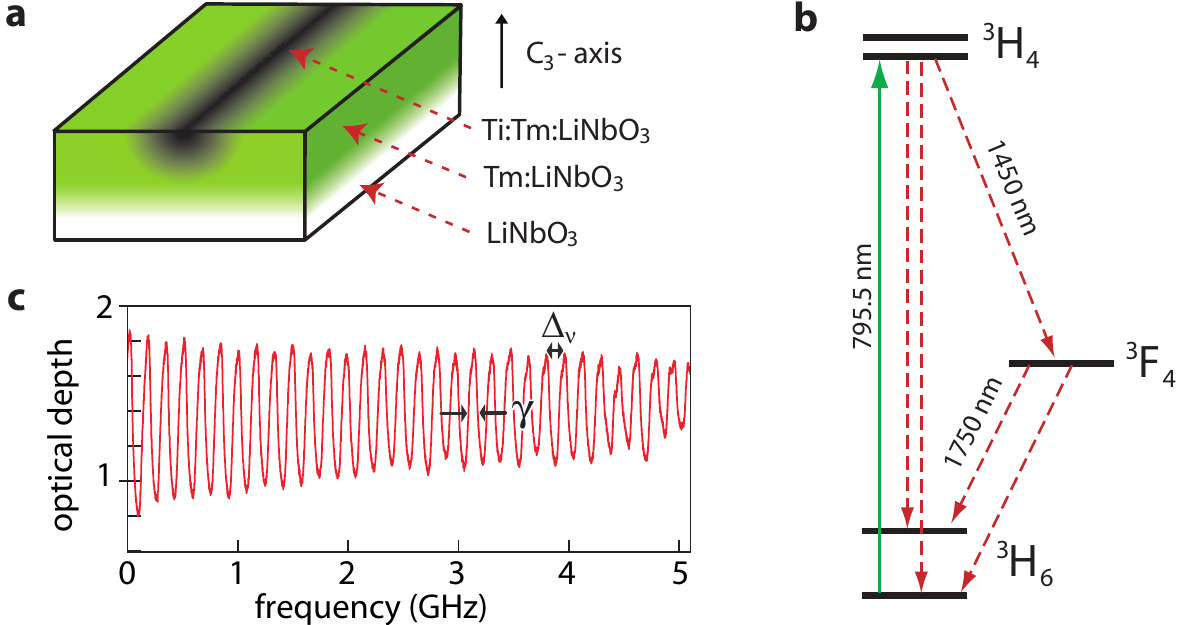}
  \caption{(color online) \textbf{a.} Waveguide geometry: 
The sample surface is first doped by indiffusing a $\approx 20$~nm thick Tm layer yielding a concentration profile of $\approx 6~\mu$m depth with $\approx 10^{20}$ ions per $\mathrm{cm}^{3}$ surface concentration. Subsequently a 3~$\mu$m wide channel waveguide is fabricated by indiffusion of a 40~nm thick vacuum-deposited Ti stripe.
AFC preparation light and single photons are coupled in and out of the waveguide with 10\% total efficiency by butt-coupling single mode fibers. \textbf{b.} Simplified energy level diagram of Tm ions: The optical coherence time of the $^3$H$_6$-$^3$H$_4$ transition at 3~K is 1.6~$\mu$s, and the radiative lifetimes of the $^3$H$_4$ and $^3$F$_4$ levels are 82~$\mu$s and 2.4~ms, respectively. A 570~G magnetic field splits the ground and excited levels into Zeeman sub-levels. The ground Zeeman level splitting is $\sim 83$~MHz, and the lifetime of the upper ground level exceeds one second. \textbf{c.} 5~GHz-bandwidth AFC: The tooth separation is $\Delta_\nu=167$~MHz, corresponding to 6~ns storage time. The line-width of the teeth is $\gamma=83$~MHz.}
  \label{fig:waveguide}
  \end{center}
\end{figure}

Suitable media in which to implement the AFC protocol are cryogenically cooled rare-earth ion doped crystals~\cite{tittel2010a,thiel2011a}. They feature inhomogeneously broadened absorption profiles, often posses long-lived atomic sub-levels that can serve as shelving levels for tailoring the AFC through persistent spectral hole burning, and generally have long coherence times on optical and spin transitions. 
We use the $^3$H$_6$-$^3$H$_4$ transition of Tm ions in a single-mode channel waveguide fabricated by Ti indiffusion into the Tm doped surface of a Z-cut LiNbO$_3$ crystal, see Fig.~\ref{fig:waveguide}a~\cite{sinclair2010a}. To tailor the desired AFC into the inhomogeneously broadened absorption profile, Tm ions with transition frequencies within the comb's troughs are optically pumped via the excited level into long-lived nuclear Zeeman levels, see Fig.~\ref{fig:waveguide}b \cite{sinclair2010a,thiel2010a}. To achieve frequency selective optical pumping we employed a linear side-band chirp technique~\cite{reibel2004a,saglamyurek2010a} that allowed us to create a 5~GHz broad grating (matching the spectral width of the 795~nm photons) with tooth spacing of 167~MHz, see Fig.~\ref{fig:waveguide}c. This corresponds to a storage time of 6~ns. After each 10~ms-long AFC preparation a 2.2~ms-long wait time allows atoms excited by the optical pumping to decay before the photon storage (see Fig.~\ref{fig:setup}b for the timing per experimental cycle). A set of micro electro-mechanical switches (MEMS) then open the channel for qubits to enter the memory, and, after recall, direct them towards the Si-APD. We assessed our memory's retrieval efficiency to be 2\%. Taking the 10~dB fibre-to-fibre coupling loss in and out of the waveguide into account, this yields an overall system efficiency of 0.2\%.

An interesting and useful aspect of photon-echo quantum memory protocols is that they provide a robust tool to manipulate time-bin qubits \cite{moiseev2004a,riedmatten2008a,hosseini2009a,moiseev2010a}. For example, using the AFC approach, any projection measurement on time-bin qubit states can be performed by superimposing two combs (double AFC) with appropriately chosen relative center frequencies and amplitudes~\cite{riedmatten2008a}. This leads to two re-emission times that can be set to differ by the  temporal mode separation of the qubit to be analyzed (1.4~ns for our experiments). Hence, as a previously absorbed photon is re-emitted by the superimposed combs, early and late temporal modes interfere, allowing the qubit state to be analyzed in the same way as is typically done with an imbalanced Mach-Zehnder interferometer~\cite{riedmatten2008a}. Double AFC recall will, however, lead to a reduction of the recall efficiency (compared to single recall).
 
\begin{figure}[tt]
\begin{center}
\includegraphics[width=0.50\columnwidth]{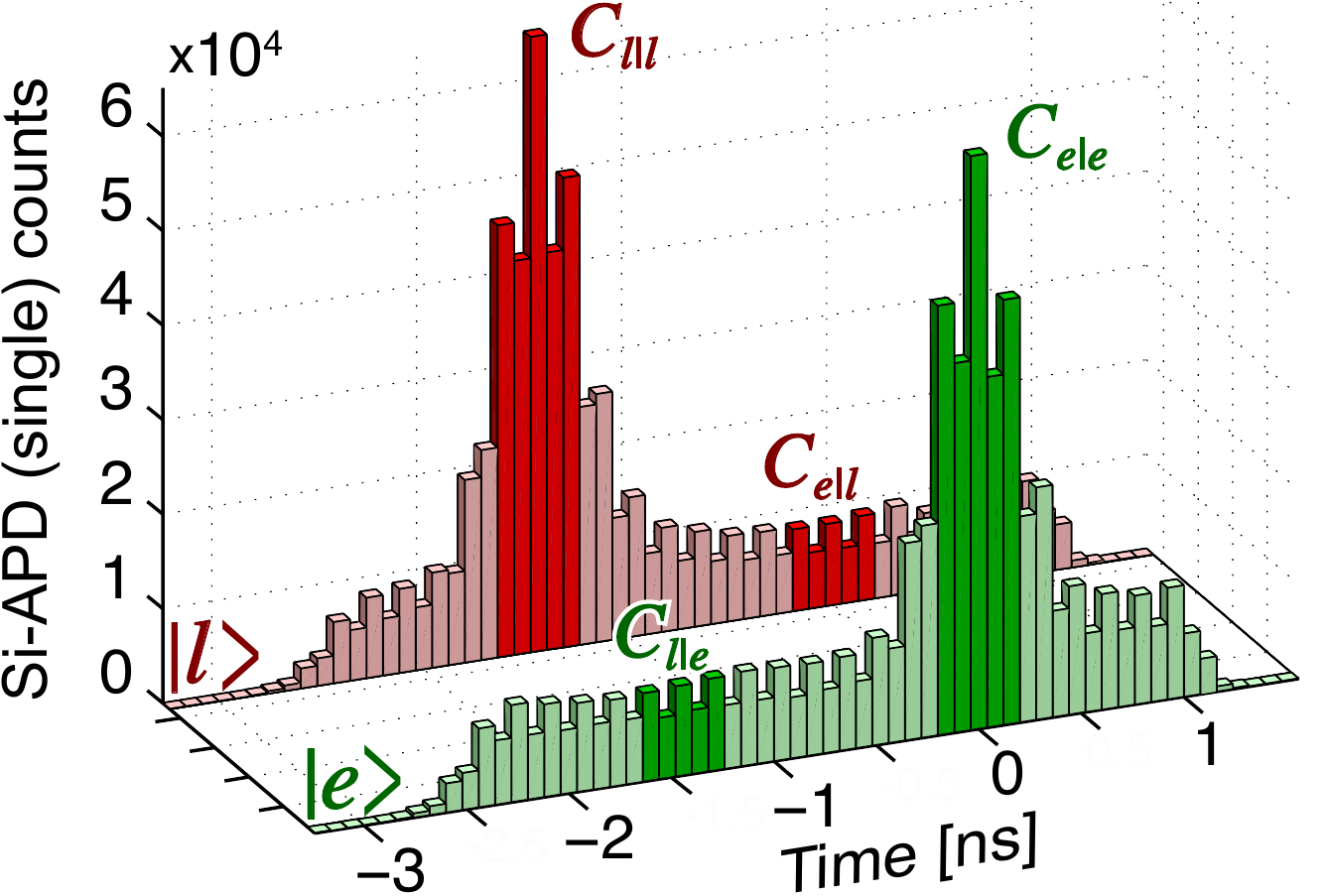}\hfill%
\includegraphics[width=0.50\columnwidth]{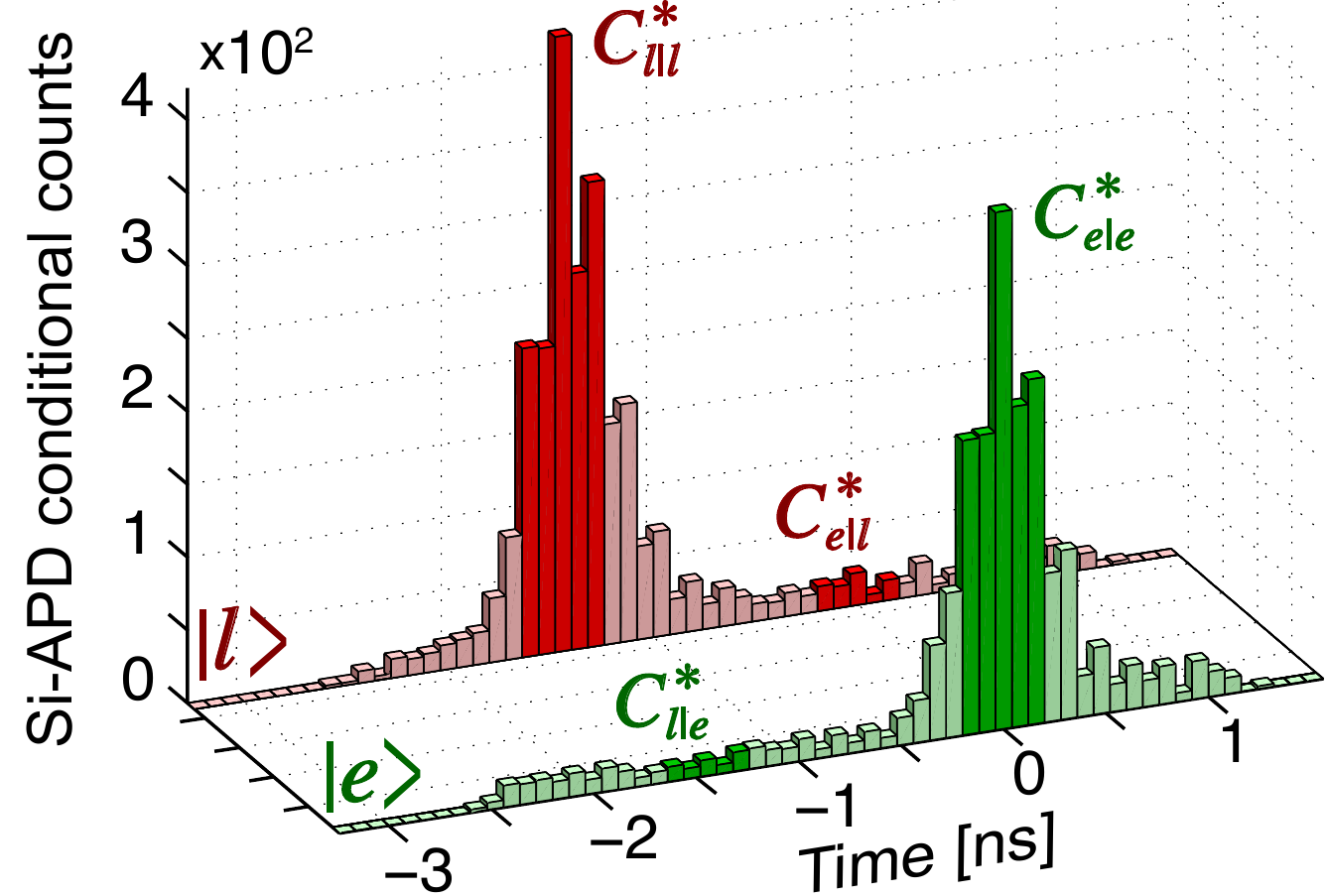}\\
  \caption{(color online) Storage of early and late time-bin qubit states in the AFC memory. The left-hand figure depicts the histograms from 180~min of single detections of the retrieved 795~nm photons prepared in early (red) and late (green) qubit states with the highlighted regions marking the relevant detection windows. The right-hand figure shows the detections conditioned on 1532~nm photons for the same states.
Without conditioning the fidelities are $\mathcal{F}_{e}=0.8652\pm0.0006$
and $\mathcal{F}_{l}=0.8376\pm0.0004$
for the storage of early and late time-bin states, respectively.
Correspondingly, with conditioning, the fidelities are
$\mathcal{F}^{*}_{e}=0.9505\pm0.0058$
and $\mathcal{F}^{*}_{l}=0.9573\pm0.0033$.
}
\label{figure_results2}
\label{fig:poledata}
\end{center}
\end{figure}

To demonstrate faithful storage and retrieval of  quantum states from the memory, we performed projection measurements with various time-bin qubits onto different bases using single (standard) and double AFC schemes as explained before. In all our measurements the average photon number per qubit was 0.1 at the output of the qubit-preparation interferometer. First we generated qubit states that occupy only early $|e\rangle$ or late $|l\rangle$ temporal modes by blocking either the long or short arm of the qubit-encoding interferometer, respectively, and then stored these states in the memory for 6~ns. Fig.~\ref{fig:poledata} (left) shows single detections (no conditioning) of the retrieved photons as a function of the time difference with respect to the START signal. The dark counts from the Si-APD reduce the signal to noise ratio (SNR) to $\sim$ 5. 
For an input state $|e\rangle$, we compute the fidelity as $\mathcal{F}_{e}=C_{e|e}/(C_{e|e}+C_{l|e})$, where, e.g., $C_{l|e}$ denotes the number of detected counts in the late time-bin given $|e\rangle$ was encoded in the qubit at the input. Similarly, we can find $\mathcal{F}_{l}$, enabling us to calculate the mean fidelity: $\mathcal{F}_{el}=(\mathcal{F}_{e}+\mathcal{F}_{l})/2=0.8514\pm0.0004$.
On the other hand, conditioning the detections of the retrieved photons on the detection of 1532~nm photons leads to a substantial increase of the SNR to $\sim 22$, as shown in Fig.~\ref{fig:poledata} (right). This yields a mean fidelity of $\mathcal{F}^{*}_{el}=0.954\pm0.015$.

\begin{figure}[tt]
\includegraphics[width=0.50\columnwidth]{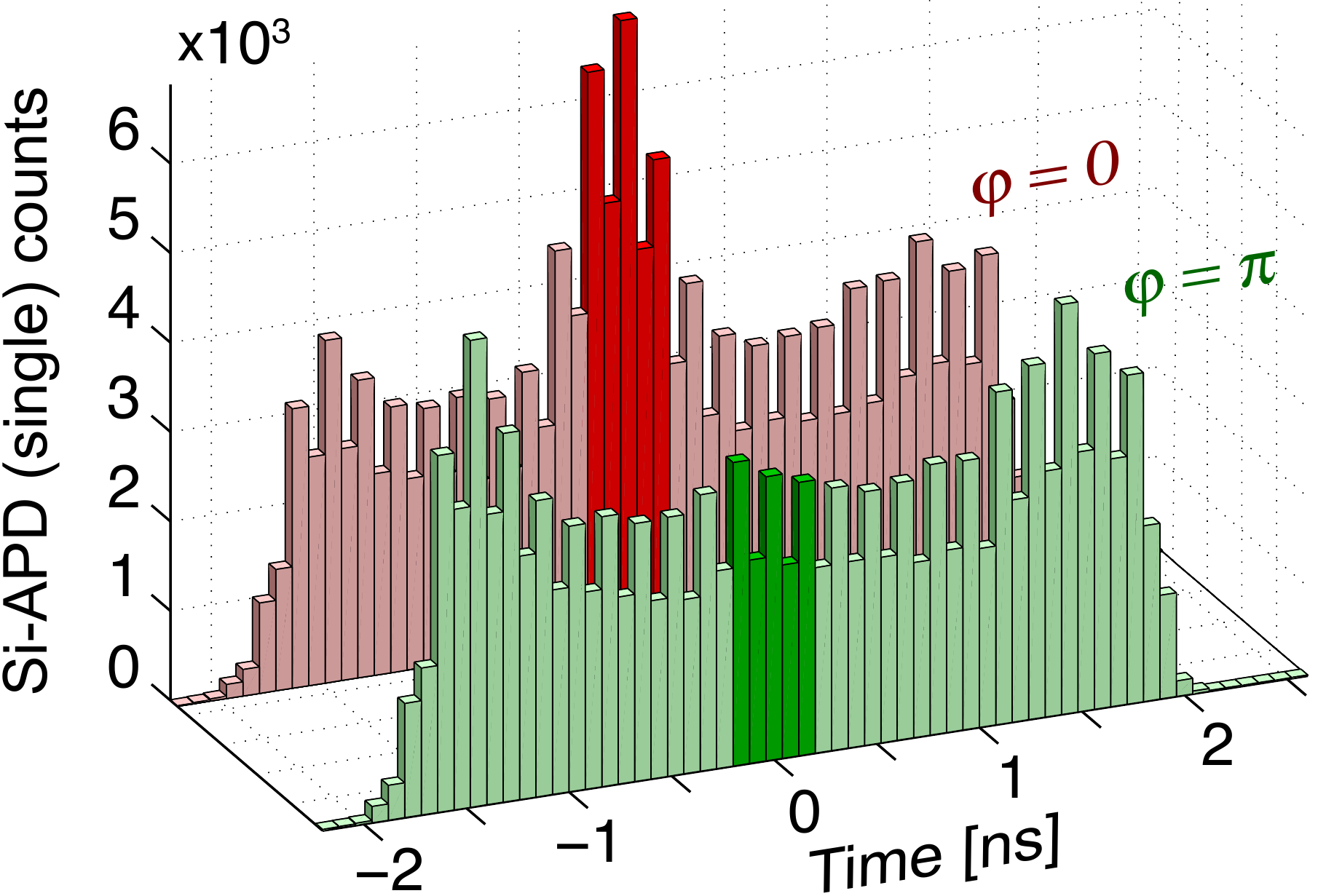}\hfill%
\includegraphics[width=0.50\columnwidth]{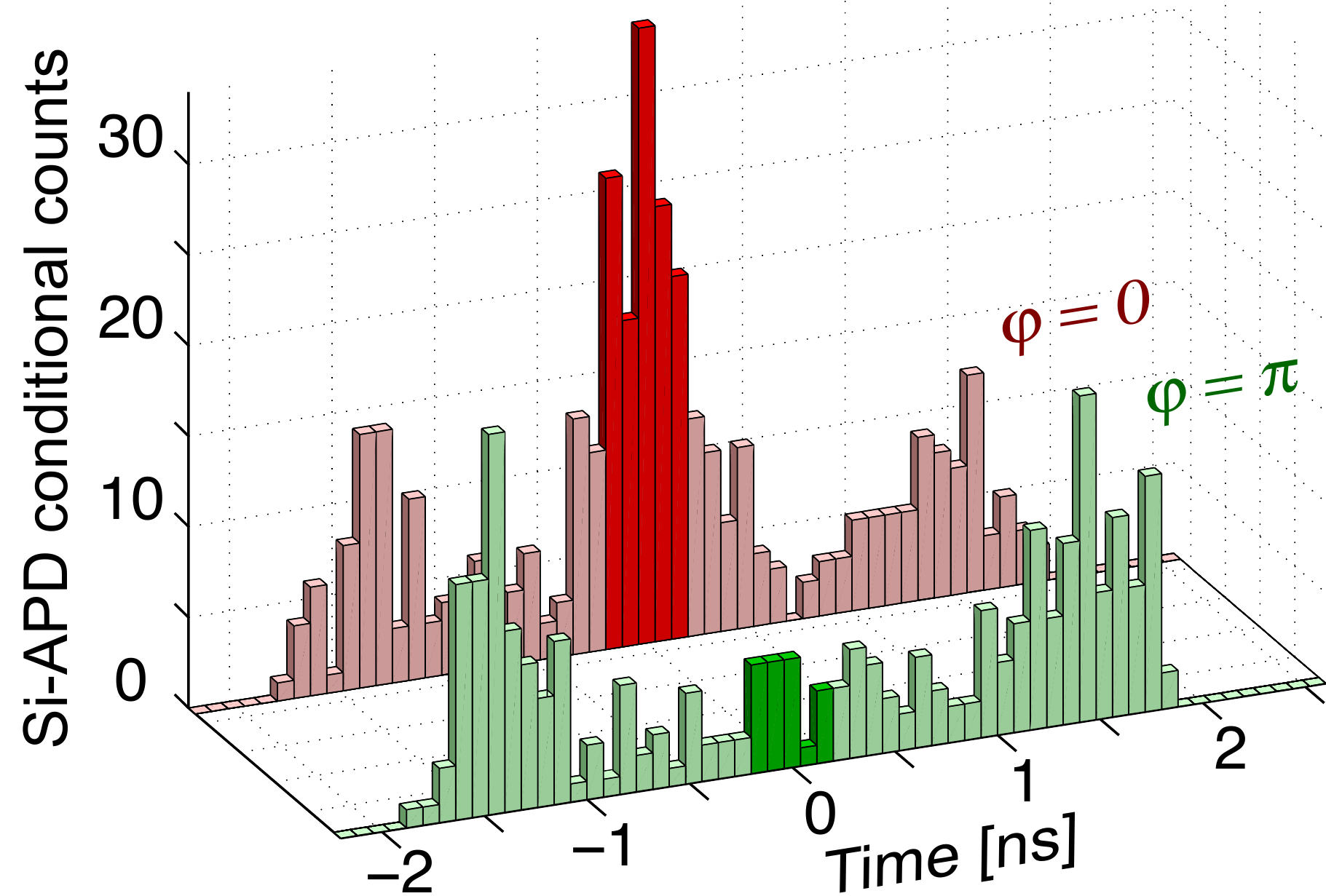}\\[10pt]
\includegraphics[width=0.47\columnwidth]{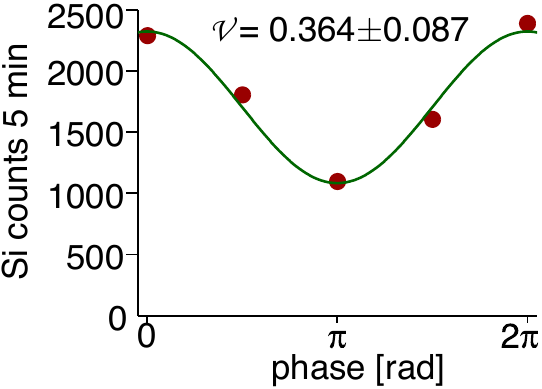}\hfill%
\includegraphics[width=0.47\columnwidth]{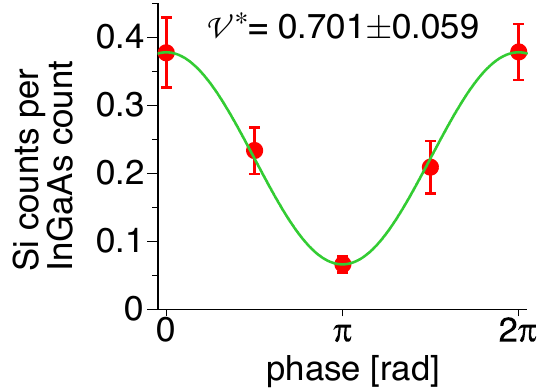}%
\caption{(color online) Retrieval of qubits created in a superposition of early and late temporal modes. The top left figure presents histograms of single detections of the retrieved 795~nm photons with AFC phase settings of zero (red) and $\pi$ (green), collected during 80~min. The top right figure shows the same histograms for conditional detections. The highlighted regions mark detection windows used to derive projection probabilities required to calculate fidelities. The lower curves show single and coincidence counts obtained for all phase settings for single detections (left) and conditional detections (right), yielding visibilities of $0.364\pm0.087$ %
and $0.701\pm0.059$%
, respectively.}
\label{fig:equatordata}
\end{figure}

Next, qubit states in an equal superposition of early and late temporal modes $\frac{1}{\sqrt{2}}(|e\rangle+e^{i\phi}|l\rangle)$ were produced with $\phi$ set to zero. Storage and projection measurements were performed using the double AFC scheme with the relative phase of the two combs (measured w.r.t. the phase introduced by the qubit-preparation interferometer) varied by $\pi/2$ increments. 
The results for single and conditional detections are given in Fig.~\ref{fig:equatordata}. The histograms show the detection statistics for zero and $\pi$ double AFC phase settings, from which we extract a SNR slightly above 1 for the single, and above 6 for the conditional detection. In the lower part of Fig.~\ref{fig:equatordata} we show the normalized counts for each projection setting for the single and conditional detections. Fitting sinusoidal curves to these we derive visibilities $\mathcal{V}$, which, in turn, yield a fidelity $\mathcal{F}=(1+\mathcal{V})/2$ for single detections of $\mathcal{F}_{\mathrm{\phi}}=0.682\pm0.020$.
For conditional detections we find a significantly larger value of $\mathcal{F}^{*}_{\mathrm{\phi}}=0.851\pm0.030$.
These figures allow establishing an average, single detection fidelity: $\overline{\mathcal{F}} \equiv (\mathcal{F}_{el}+2\mathcal{F}_{\mathrm{\phi}})/3=0.738\pm0.029$.
This violates the quantum classical bound~\cite{massar1995a} of $\sim$ 0.667, thus verifying that our memory outperforms any classical storage protocol. However, it is below the bound of $\sim$ 0.833 for an optimal universal quantum cloner~\cite{buzek1996a}.
Harnessing the conditional detection we find $\overline{\mathcal{F}^{*}}=0.885\pm0.020$.
This beats the quantum-classical bound by 10 standard deviations and also violates the optimal universal quantum cloner bound by 2.5 standard deviations.

To conclude, we have demonstrated storage, retrieval, and conditional detection of different time-bin qubit states using a solid-state Ti:Tm:LiNbO3 waveguide quantum memory with average fidelity $\overline{\mathcal{F}^{*}}=0.885\pm0.020$, which  exceeds the relevant classical bounds. Operating the memory in a heralded fashion is readily achievable with high-rate APDs that have recently become commercially available. Despite our memory device's current limitations, namely efficiency, storage time, and preset recall time, the high fidelity and the wide spectral acceptance makes our approach promising for future quantum communication schemes and quantum networks.
The LiNbO$_3$ host crystal and the waveguide structure have potential advantages in quantum memory applications such as fast electric field control of collective atomic phase evolution and, due to the resemblance with building blocks of classical integrated optical devices~\cite{sohler2008a}, it holds promise for simple integration with existing information technology.
Furthermore, the ability to perform projection measurements using a photon-echo memory provides a simple and robust tool that might find use in other applications of quantum information processing. 

\acknowledgments     
We thank C. La Mela and T. Chaneli\`{e}re for helping in the initial stages of this work, V. Kiselyov for technical support, and NSERC, GDC, iCORE (now part of AITF), QuantumWorks, CFI and AET for financial support. D.O. thanks the Carlsberg Foundation and F.B. thanks FQRNT for support.


\end{document}